# When Your Friends Become Sellers: An Empirical Study of Social Commerce Site Beidian

Hancheng Cao[*1], Zhilong Chen[*2], Fengli Xu[2], Tao Wang[3], Yujian Xu[4], Lianglun Zhang[4], Yong Li[†2]

[1]Department of Computer Science, Stanford University
[2]Beijing National Research Center for Information Science and Technology,
Department of Electronic Engineering, Tsinghua University
[3]Grenoble Ecole de Management, [4]Beibei Group
hanchcao@cs.stanford.edu, {chen-zl16,xfl15}@mails.tsinghua.edu.cn, tao.wang@grenoble-em.com,
{yujian.xu,lianglun.zhang}@beibei.com, liyong07@tsinghua.edu.cn[‡]

## Abstract

Past few years have witnessed the emergence and phenomenal success of strong-tie based social commerce. Embedded in social networking sites, these E-Commerce platforms transform ordinary people into sellers, where they advertise and sell products to their friends and family in online social networks. These sites can acquire millions of users within a short time, and are growing fast at an accelerated rate. However, little is known about how these social commerce develop as a blend of social relationship and economic transactions. In this paper we present the first measurement study on the full-scale data of Beidian, one of the fastest growing social commerce sites in China, which involves 11.8 million users. We first analyzed the topological structure of the Beidian platform and highlighted its decentralized nature. We then studied the site's rapid growth and its growth mechanism via invitation cascade. Finally, we investigated purchasing behavior on Beidian, where we focused on user proximity and loyalty, which contributes to the site's high conversion rate. As the consequences of interactions between strong ties and economic logics, emerging social commerce demonstrates significant property deviations from all known social networks and E-Commerce in terms of network structure, dynamics and user behavior. To the best of our knowledge, this work is the first quantitative study on the network characteristics and dynamics of emerging social commerce platforms.

## Introduction

Social commerce, is known as the act of using social media to promote online buying and selling of products and services (Liang and Turban 2011; Zhou, Zhang, and Zimmermann 2013; Curty and Zhang 2011). Dating back to 2005, there have been numerous attempts to enable social commerce, including adding social media features in E-Commerce sites (e.g., Amazon customer reviews, Groupon (Hughes and Beukes 2012)), or using existing social media to do marketing (e.g., Facebook-based F-commerce, Twitter-based T-commerce (Culnan, McHugh, and Zubillaga 2010)). While traditional social commerce heavily relies on key opinion leaders (KOL) such as celebrities and brands (Yadav et al. 2013), a recent trend shows that leveraging the social ties of ordinary people to do marketing can be highly impactful. In China several social commerce sites (e.g., Pinduoduo[1], Yunji[2], Beidian[3]) have been demonstrating how marketing through instant messaging app WeChat (with 1 billion monthly active users) can turn family members and friends into repeated and loyal consumers (Meeker 2017; 2018). Typically, users of social commerce platforms share product information in WeChat group with families and friends either through direct messaging or posting entries visible to them. In this way, information cascades and new users pour in. Pinduoduo, for instance, acquired over 200 million users in less than three years and the daily order volume ranks second in mainland China next to Taobao, posing considerable threat to traditional e-commerce giants like Alibaba, Amazon and JD [4].

How to explain the speed and scale of the growth of strong-tie based social commerce? Do user behavior on such emerging social commerce sites differ from traditional E-Commerce? If so, what explains the differences? In this paper, we leverage a full-scale dataset of Beidian, one of Chinese most popular WeChat-based social commerce platforms, to shed light on these questions.

Complementing qualitative/theoretical studies in social commerce, we rooted our analysis in network science and empirical data analysis. More specifically, we tackled how the growth of strong-tie based social commerce platforms is grounded in the interaction between social relationship and economic logics. First, we introduced the topological structure of the overall Beidian network, demonstrating that the network is decentralized and consists primarily of ordinary people with limited social influence. We then illustrated invitation mechanism behind Beidian's rapid growth, revealing the distinctive nature of deeper and larger invitation cascades on Beidian compared to previously studied networks. Finally, we investigated user purchasing behavior patterns on Beidian. Findings show that Beidian users within a seller community show high degree of proxim-ity and loyalty. Both factors correlate with the site's high conversion rate. As the confluence of social closeness and economic transactions, emerging social commerce demonstrates significant property deviations from all known social networks and E-Commerce in terms of network structure, dynamics and user behavior. To the best of our knowledge, this work is the first quantitative measurement study on the network properties and dynamics of strong-tie based social commerce platform.

---

[1]https://www.pinduoduo.com/
[2]https://www.yunjiweidian.com/
[3]https://www.beidian.com/
[4]https://www.forbes.com/sites/alexfang/2018/07/26/ipo-of-chinese-e-commerce-firm-pinduoduo-mints-new-young-billionaire/50da2a734024

This paper is organized as follows. We first review related works. Then we introduce Beidian's operation and provide detailed description of our studied dataset. Next we study the structure of Beidian network, analyze its growth and invitation mechanism via cascading tree, and demonstrate Beidian's high conversion rate as well as user purchasing pattern. We conclude by discussing implications of our empirical findings and proposing the novel characteristic of intimacy-based social commerce sites.

## Related Work

We summarize the most related works in three aspects: measurement of social network & social media, user behavior analysis on E-Commerce sites, and social commerce studies.

**Measurement of Social Network & Social Media.** Measurement works aim to understand the fundamental characteristics of online social network and social media sites. (Kwak et al. 2010) crawled the entire Twitter site and studied its network structure, users and trending topics. (Wattenhofer, Wattenhofer, and Zhu 2012) examined the social network aspect of YouTube using its subscription graph, comment graph and video content corpus. Other studies investigated Flicker (Mislove et al. 2007), Microsoft Messenger (Leskovec and Horvitz 2008), Foursquare (Noulas et al. 2011), Instagram (Hu et al. 2014), etc.

In social network analysis, the effect of network growth, information cascades and homophily have received significant attention. For instance, (Zang, Cui, and Faloutsos 2016) studied the growth of several online social networks and proposed the network growth model NETTIDE. (Anderson et al. 2015) studied the sign-up cascade of LinkedIn network, which resulted in the rapid growth of LinkedIn. Homophily, the effect of 'birds of a feather flock together', has been found prevailing in most social networks (McPherson, Smith-Lovin, and Cook 2001). It has also been shown that user purchase intent can be identified and predicted using social media posts (Zhang and Pennacchiotti 2013; Gupta et al. 2014).

Building on and extending previous works on social network analysis, our paper focus on the measurement of a newly emergent social commerce site based on real life social network.

**User Behavior Analysis on E-Commerce sites.** Previous works studying user behavior on E-Commerce sites focused on quantitative analysis of log data. (Lo, Frankowski, and Leskovec 2016; Zeng et al. 2019) analyzed the purchasing behavior of 3 million Pinterest users and examined how both short-term and long-term signals reflect user purchasing intent. (Kooti et al. 2016) analyzed user behavior patterns on E-Commerce using email confirmation data and demographic information. An interesting work (Guo, Wang, and Leskovec 2011) analyzed the role of instant messaging tool integrated in Chinese E-Commerce site Taobao, and measured its influence on information passing and purchasing decision. Meanwhile, word-of-mouth recommendation and viral marketing (Richardson and Domingos 2002; Matsuo and Yamamoto 2007) have been popular practices in E-Commerce. (Leskovec, Adamic, and Huberman 2007; Leskovec, Singh, and Kleinberg 2006) studied a large person-to-person product recommendation network and revealed the effectiveness of viral marketing. Furthermore, (Aral and Walker 2011) and (Aral, Muchnik, and Sundararajan 2013) studied viral product design strategies for E-Commerce and simulated social contagion in the presence of homophily.

While prior works focused on E-Commerce formed primarily by strangers (i.e., seller and buyer generally do not know each other) (Guo, Wang, and Leskovec 2011; Ye et al. 2012), our work investigates the recent trend of marketing over strong social ties, and highlights the interplay between social network and E-Commerce transaction network, which potentially opens up multiple new directions in E-Commerce research.

**Social Commerce.** Social commerce refers to any act of using social media to promote online buying and selling of products and services (Liang and Turban 2011; Zhou, Zhang, and Zimmermann 2013; Curty and Zhang 2011). Social commerce varies in forms (Curty and Zhang 2011) and can be categorized into social network driven platforms (e.g., F-commerce, T-commerce), group buying platforms (e.g., Groupon), peer-to-peer sales platforms (e.g.EBay), and peer recommendation platforms (e.g. Amazon). Earlier works in sociology and economics literature studied the embeddedness of economic action within social structure (Granovetter 1985), and the interplay between market organization and trading relationships (Geertz 1978; Weisbuch et al. 1997), which lay the theoretical foundation of social commerce research. (Wang and Zhang 2012) chronologically traced the evolutionary patterns of social commerce, and analyzed its development from the dimensions of people, management, technology and information. (Stephen and Toubia 2010) stated that social commerce's value lies primarily in making shops more accessible to users and greatly benefit people whose accessibility is enhanced by the network. (Kim and Park 2013) and (Shin 2013) found that characteristics of social commerce have significant impacts on users' trust and in turn affects users' purchasing behavior, while (Holtz, MacLean, and Aral 2017) studied the role of networked social signals in generating trust within peer-to-peer platforms. (Bolton, Katok, and Ockenfels 2004) and (Cai et al. 2014) further studied the influence of electronic reputation mechanism and its effect on trust/trustworthiness in social commerce. Meanwhile, (Chen et al. 2018) proposed a bilateral-attention LSTM to detect social commerce sellers through WeChat posts. However, existing research on social commerce are either qualitative/theory driven (Wang and Zhang 2012), or analyze social commerce of different mechanisms (Lu, Fan, and Zhou 2016; Holtz, MacLean, and Aral 2017). This paper adds to social commerce studies by conducting the first quantitative measurement on a full-scale invitation/transaction dataset from one of the largest and fast growing strong-tie based social commerce sites.

## Background and Dataset

Beidian is one of the largest and fastest growing WeChat-based social commerce sites in China. Founded in August 2017, Beidian has acquired over 44.85 million users by the end of 2018 [5]. In contrast to other social commerce platforms, Beidian adopts a strong-tie based business model: The Beidian platform is built on top of WeChat and directly leverages existing WeChat kinship and friendship relations for marketing (Wang, Li, and Tang 2015; Church and De Oliveira 2013), i.e., Beidian users mostly sell and buy with someone they know well in real life instead of strangers.

Two types of actors are present on Beidian: sellers and buyers. A seller acts like an agent - he/she does not hold any inventory, but selects and advertises certain Beidian products through sharing links with their friends in WeChat groups (communities) or post links on their friends-visible status (WeChat moment) via Beidian's app, as shown in Fig. 1(a). By clicking the product link on WeChat shared by their friends, potential buyers are directed to the product's webpage on Beidian, where they can view detailed product information (e.g., price, description, popularity, etc.) and make a purchase (Fig. 1(b)). If a seller's advertisement successfully leads to purchase on Beidian, Beidian will share part of the revenue with the seller. Unlike traditional E-Commerce, anyone can become a seller on Beidian once they pay a small fee (roughly $60). To shop on Beidian, one can only register and become a valid seller through the invitation link from an existing Beidian user (seller/buyer), except for the very first few 'seed users'. Therefore, for a typical Beidian user, he/she first joins the platform via invitation sent by his/her WeChat friends, then purchases items promoted in friends' WeChat group/moment. If he/she later decides to submit a fee and becomes a seller himself/herself in hope of getting commission from Beidian, he/she would choose some Beidian products (which are potentially of interest to his/her WeChat friends) and make efforts to advertise them by sharing product links on WeChat. He/she could also invite new Beidian users within his/her social circle so as to promote sales. It is important to note that Beidian differs intrinsically from multi-level marketing (Bloch 1996). Whereas in multi-level marketing, cascading recruitment of sellers form selling pyramid and profits are split between sellers and up line distributors in different levels (Nat and Keep 2002), profits generated on Beidian are only shared between sellers and the platform without splitting with other intermediaries.

We obtained anonymized Beidian user behavioral data through a research collaboration with Beibei Group (Beidian's parent company). Due to Beidian's mechanism, our dataset is therefore composed of two basic components: user invitation records and purchase records.

---

[5] http://tech.caijing.com.cn/20190107/4553034.shtml

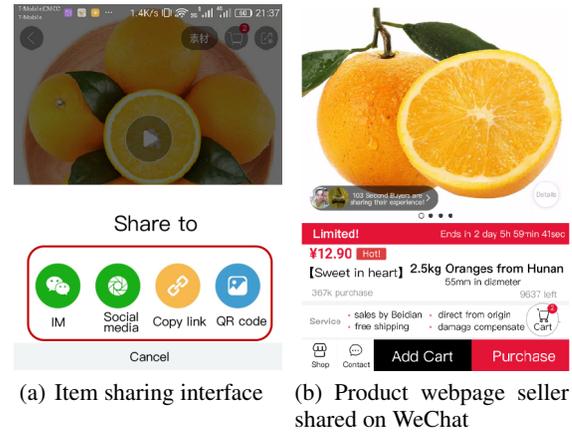

(a) Item sharing interface  (b) Product webpage seller shared on WeChat

Figure 1: Interface of Beidian.

**Invitation Records.** This part of data include the complete invitation records of Beidian users from its launch to June 4th, 2018. A total of 11,853,205 users joined Beidian during the period. Typically, existing members share sign-up invitations with their family members and friends via WeChat. If they succeed in referring a new member, they will get bonus from Beidian, thus existing members are usually highly motivated in the advertising campaign. In our dataset, the exact sign-up timestamp and the inviter of each user are recorded.

**Purchase Records.** This part of data include the full-scale purchase history of Beidian users from May 31st, 2018 to November 27th, 2018 and the complete behavioral records of Beidian users from July 31st, 2018 to November 27th, 2018 respectively. 2,962,880 users made 15,786,527 purchases within the six months' period and all users' complete behaviors (e.g., browsing, cart adding, etc) on Beidian in the last four months were recorded. Items on Beidian are grouped into 12 categories: paper & household cleaning, dietary supplements, baby clothes, snacks, fruits & vegetables, milk powder & diapers & baby food, toys, cosmetics & skin care, grains & cooking oils & drinks, household supplies, personal care and clothes & shoes & bags.

*Ethical Considerations.* We took careful steps to address privacy issues regarding the sharing and mining of user behavioral data. Firstly, consent for research studies is included in the Terms of Service for Beidian. Secondly, pre-processing is conducted before we obtained the data, through which user privacy is protected. All user identifiers have been replaced with secure hashcodes to improve anonymity. Furthermore, our local university institutional board has reviewed and approved our research protocol. Finally, all data is stored in a secure off-line server, with access limited to only authorized members of the research team bound by strict non-disclosure agreements.

## Overview: A Decentralized Network

To understand the structure of strong-tie based social commerce, we first present a topological analysis of Beidian network. Specifically, we study the number of new users each inviter invited to Beidian and the number of shoppers each seller sold items to through constructing invitation/purchase graphs, where nodes depict Beidian users and edges represent invitation/purchase relationship.

The out degree distributions of the invitation graph and the purchase graph are delineated in Fig. 2(a) and Fig. 2(b) respectively. From the figures, we can clearly observe that most of the inviters and sellers do not have a large out degree, indicating that most users only invite a few people into the network, while most sellers sell products to a few buyers. To be specific, 48.3% of the inviters' out degree does not exceed 10, and the vast majority of the inviters (96.5%) show out degree less than 100. Similar patterns are spotted in the out degree distribution of the purchase network, i.e., 53.4% of all sellers have no more than 10 customers, and 96.7% of all sellers sell their products to 100 customers or fewer. Moreover, both out degree distributions greatly deviate from power law commonly observed in various online social networks (Muchnik et al. 2013), or fatter distribution represented by Twitter (Kwak et al. 2010). Instead, the node distribution of Beidian skews greatly towards small value, indicating lack of high-degree nodes, or key opinion leaders, in this network. Fig. 3(a) and Fig. 3(b) further show the proportion of invitation/purchase activities by degree of inviter/seller. 72.1% of all successful invitations are sent by inviters with out degree less than 100, while 73.2% of all purchases are made under sellers with fewer than 100 buyers, both of which indicates that the majority of invitation/purchase activities are carried out by users of limited social influence.

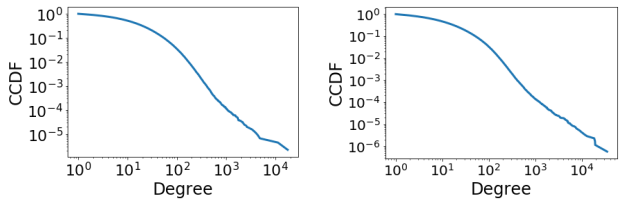

(a) Degree distribution of inviters  (b) Degree distribution of sellers

Figure 2: Degree distribution of inviters/sellers.

Therefore, Beidian network is highly decentralized. These observations provide first evidence of the intimate nature of buyer-seller relationships embedded in real life social network. Unlike most social media based commerce such as Twitter commerce and Facebook commerce, where certain brands or public pages followed by millions users established themselves as key opinion leaders (McCormick 2016), the Beidian network does not involve 'super nodes' and is made up primarily of ordinary people, who can only exert 'local influence' on their direct neighbors. Due to the nature of WeChat, this decentralized network resembles people's real life social relations, such as family members and

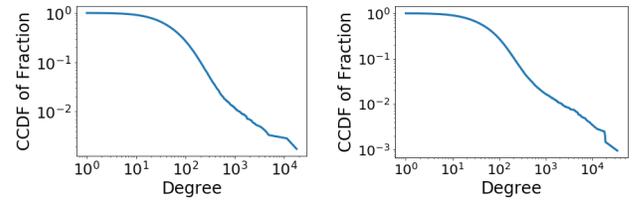

(a) Proportion of invitation by degree of inviter  (b) Proportion of purchase by degree of seller

Figure 3: Proportion of invitation/purchase activities by degree of inviter/seller.

friends. Moreover, such network covers a wide range of demographic features, including age, gender, economic status or interests. Personal ties are mapped onto E-Commerce and invitations to buy products can be interpreted as entitlement, obligation, or gift. Thus, the intertwining of economic transactions and strong social relationship could potentially help Beidian grow and expand differently from conventional E-Commerce sites. To shed light on this, we now turn to the dynamics of Beidian, where we focus on two most important activities for E-commerce: enrollment of new members, and interaction of existing users (i.e., marketing & purhcasing). In particular, we show how a) Beidian grows through invitation cascades and b) purchases are shaped by social closeness, contributing to the site's high conversion rate.

## Growth via Invitation

In this section, we present a measurement analysis on Beidian's growth patterns.

### Growth Speed

Beidian is featured by its rapid growth of user and transaction numbers. As illustrated in Fig. 4, within the first 10 months, Beidian has grown into an E-Commerce giant with nearly 12 million users. In the recent six months, 15,786,527 purchases has been made. Fig. 5 further shows the number of new users joining the platform and purchases made daily on Beidian. We can also observe that there is a clear trend that the daily new users and purchases increase as time goes by (the peak around 100 days since Beidian's launch is due to the promotion at Chinese Double 11 Festival), indicating that the growth of Beidian is accelerating.

Why did Beidian grow so fast? As a new user has to first receive an invitation link from one existing Beidian member so as to become a valid shopper before making a purchase, we now investigate the site's unique invitation cascade patterns, including the depth, size and virality of Beidian's invitation cascades through cascading tree.

### Invitation Cascade

To study Beidian's invitation cascade, we analyze the characteristics of Beidian's invitation cascade tree, where each node represents a Beidian user and edges between users indicate invitation relationship. As Beidian starts its invitation-only user recruitment from a number of seed users, the Beid-

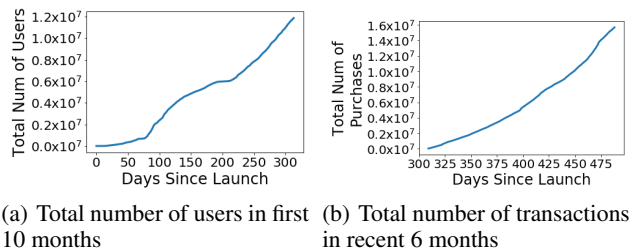

(a) Total number of users in first 10 months

(b) Total number of transactions in recent 6 months

Figure 4: Beidian's growth over time.

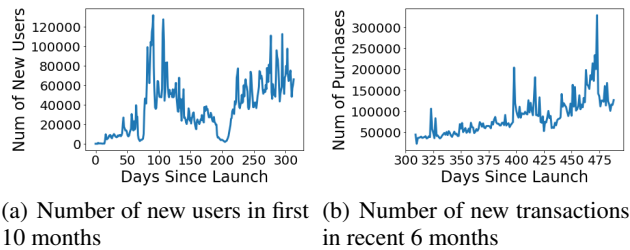

(a) Number of new users in first 10 months

(b) Number of new transactions in recent 6 months

Figure 5: Beidian's growth rate over time.

ian invitation cascades involve several trees. Fig. 6 illustrates snapshots of a cascade tree at different timestamp originated from the same root user. As shown in Fig. 6(b), the root user successfully invited 2 new users into the Beidian network on the 3rd day after himself got invited. Within 7 days, the invitees of the root user started to invite their friends into the network, forming the third layer of the cascade tree (Fig. 6(c)). The cascade tree now resides a total of 12 members. By the end of the first month since the root user joined Beidian, the fourth layer of the cascade tree has appeared, adding another 21 users to the network (Fig. 6(d)). Note that for the purpose of visualization, we selected a root user with relatively small out degrees and shallow depths, while in fact most invitation cascades on Beidian are deeper and larger, and as a consequence, resulted in even more descendants.

**Cascade Structural Patterns** Fig. 7 illustrates the normalized distribution of users' adoption depth, i.e., the number of steps between a user and the root of the cascade trees he/she is located. Different from prior works (Anderson et al. 2015; Goel et al. 2015; Leskovec et al. 2007), not only is the cascade trees of Beidian deeper, but more users are at a greater depth as well. Specifically, 71.0% of all users in Beidian invitation network are at depth 5 or beyond, with 22.6% of users lying at positions with depth more than 10. Invitation cascade can even happen at depth 24. In comparison, in prior studies, fewer than 30% of the invitations can reach depth 5. Thus, the invitations of each individual user exhibit impact much more far-reaching than their prior counterparts.

To better illustrate the virality of the invitation network, we examine the fraction of adoptions residing in the invitation cascades. Fig. 7 and Fig. 8 exhibit the fraction of users in trees of various depths and sizes. Substantial differences can once again be observed between Beidian invitation cascades and those cascades previously analyzed. For example, on LinkedIn, only 10% of users belong to cascades with size at least 10,000, but on Beidian this number can be high up to 79.9%, and most surprisingly, 64.9% of users constitute a huge cascade tree with more than 7,500,000 members. In other words, Beidian's growth is primarily the result of a few large cascades, rather than a great many small ones.

In terms of the maximum depth of the trees, distinctions could also be clearly observed. For instance, those at trees with a maximum depth of 6 and more take up 83.4% of all users while this portion is between 0.1% and 6% in earlier studies. More than 78.6% of all users reside in trees whose maximum depths exceed 10 and 64.9% of users contribute to the formation of the largest component with a maximum depth of 24.

To quantify the structural patterns of the invitation cascade, we measure the structural virality of cascades across different size groups. Following the methods of previous works (Goel et al. 2015; Anderson et al. 2015), we use the Wiener index, which gauges the average distance of paths between nodes in a tree, as the demonstration of structural virality and only those cascades with more than 100 nodes are considered. We illustrate the results in Fig. 10. The results show that the structural virality of a cascade positively correlates with its size, indicating that cascade gets deeper and relatively narrower as the size of cascade grows, which differs from Twitter cascades (Goel et al. 2015), while in line with LinkedIn cascades (Anderson et al. 2015). However, we also note that Beidian's invitation cascades demonstrate property deviations from their counterpart on LinkedIn. To be more specific, the average Wiener index of cascades with size more than 100,000 surpasses 15 on LinkedIn, but on Beidian it never exceeds 15. One possible explanation is that Beidian's users are financially motivated, so they are more likely to broadcast the invitation to as many people around them as possible (as illustrated in Fig. 6), which can lower the Wiener index to some extent but does not change its viral nature.

**Invitation Temporal Patterns** Next, we study temporal patterns of invitation cascades, where we look at how invitation behavior of new users deviates from older users. Specifically, we measure the number of successful invitations in May 2018 by users who joined Beidian at different time points, as illustrated in Fig. 11. While new Beidian users who joined the platform in May 2018 successfully enrolled 7.36 users on average, older users who joined in Aug 2017 only succeeded in inviting 0.78 users. This suggests a declining success rate of invitation over time, making users less likely to grow into a 'super node'. While decreasing motivation of individual users in inviting new users provides some rationales, we argue that the intimate nature of social network on WeChat is of additional explanatory value: there exists an upper bound to the number of people with whom one can maintain stable social relationships (also known as Dunbar's number (Dunbar 1992)). Thus, after the initial success of introducing people in one's close circle to Beidian, it become increasingly difficult for him/her to invite new members, hindering further broadening individual seller influence, which is reflected in Beidian's decentralized

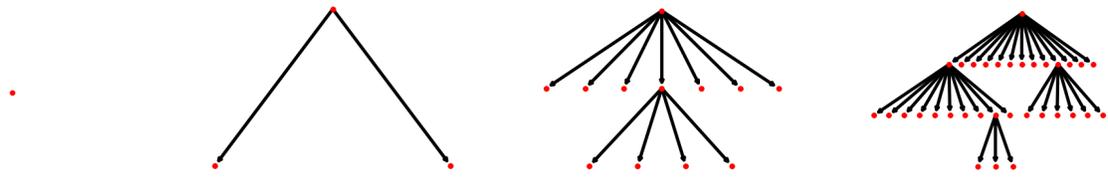

(a) Day 0: the root user joined Beidian network.
(b) 3 days after the root user joined Beidian network.
(c) 7 days after the root user joined Beidian network.
(d) A month after the root user joined Beidian network.

Figure 6: An invitation cascade on Beidian originated from a root user. Each red point represents a Beidian user, while edge indicates invitation relationship.

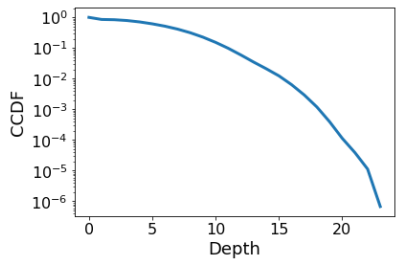

Figure 7: Distribution of adoption depth.

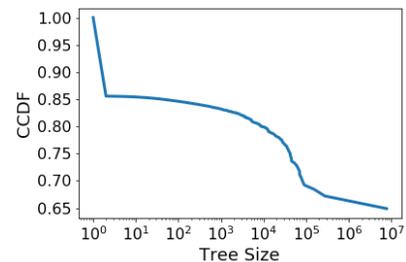

Figure 9: Fraction of users in trees of specific size.

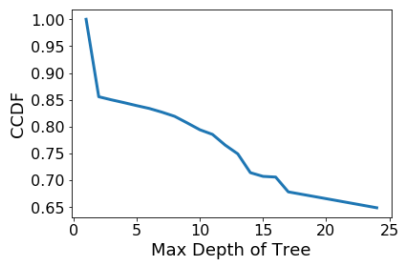

Figure 8: Fraction of users in trees of specific depth.

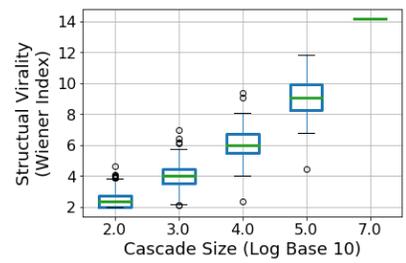

Figure 10: Structural virality of cascades with specific size.

characteristic discussed in section 4.

Taken together, Beidian cascade is viral in nature. In comparison with prior works, Beidian's invitation cascade occurs much deeper and larger. Different from traditional "point to point" invitation/recommendation in viral marketing, existing users share invitation links in WeChat groups, thus demonstrating 'point to group' phenomenon. Probably due to this invitation-only user enrollment policy conducted on intimate social circle, mingled with financial rewards which motivate participation, Beidian, as well as other strong tie based social commerce platforms, rapidly grows after its launch.

## Economic Transactions over Strong Tie

Apart from the unique mechanism of enrolling new members, strong-tie based social commerce also deviates from its prior counterparts on seller-buyer relationship. As discussed above, the seller-buyer relationship on Beidian typically represents close relationship in real life (i.e., family, co-workers, friends etc.). In this section we analyze how such social closeness affects marketing/purchasing activities on Beidian.

### High Conversion Rate

As an important performance metric of marketing strategies, conversion rate indicates users' motivation to make a purchase. We study Beidian's conversion rate, which is defined as the percentage of visits (clicks of Beidian product links shared on WeChat) that results in purchases (Moe and Fader 2004). Overall, we find that Beidian demonstrates high conversion rate. For the entire site, Beidian achieves 7.33% of conversion rate, whereas conversion rate on traditional e-commerce sites rarely exceeds 5% (Moe and Fader 2004; Wolfgang 2019). In terms of mean conversion rate under each seller, the conversion rate achieves 5.6%. Speaking of individual buyers, 35.3% of all buyers make at least one pur-chase after visiting links shared by Beidian sellers. 28.4% of the buyers show conversion rate greater than 5%, as illustrated in Fig. 12. On average, the mean conversion rate for individual Beidian buyer is 7.3%, and the mean conversion rate is 20.8% for buyers with at least one purchase, in contrast to 3.36% user conversion rate on other sites (Wolfgang 2019).

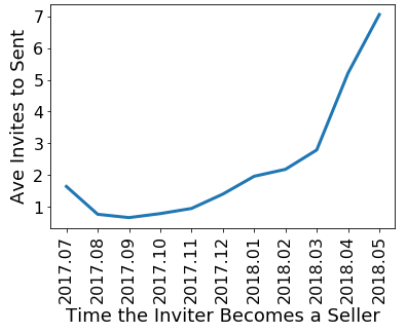

Figure 11: Average number of successful invitations in May 2018 by users joined at different time.

We further study Beidian's item conversion rate. Fig. 13 demonstrates the average item conversion rate across different product categories. We observe that conversion rate varies conspicuously from category to category. Although most categories share a conversion rate of approximately 6% to 8%, conversion rate on certain categories can be remarkably high. For example, for fruits & and vegetables, every 6-7 visits to an item link on Beidian can lead to a purchase, resulting in an average conversion rate of 16.3%. Fig. 14 further illustrates the average item conversion rate as a function of price. As can be observed, cheaper items show much higher conversion rate than more expensive ones, indicating buyers are less hesitant when buying items of lower cost.

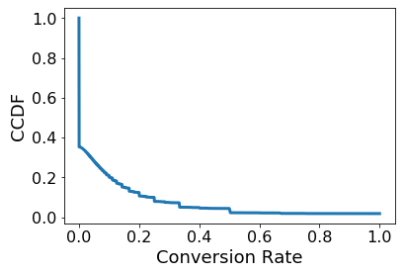

Figure 12: Distribution of user conversion rate.

## User Proximity

As Beidian network is based on social closeness, proximity analysis (Boschma 2005; Agrawal, Kapur, and McHale 2008) can help better understand the similarity across individual users, which would potentially reflect in their purchase behavior.

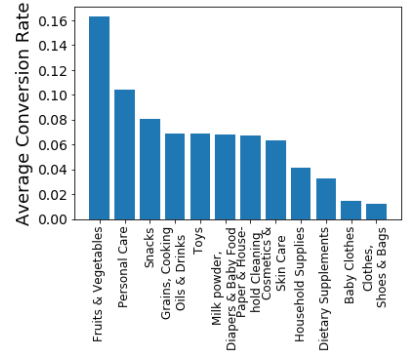

Figure 13: Conversion rate for items of different categories.

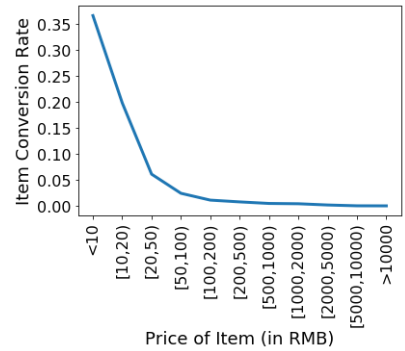

Figure 14: Conversion rate for items of different prices.

**Proximity of Users**  As we argue that the buyer-seller relationships on Beidian resemble user social network in real life, we study physical and social proximity among users using their social demographic characteristics. In our dataset, the available user social demographic characteristics include users' delivery locations at both city and province level. We approximated the economic status of the city through Chinese city tier system.

Specifically, we examine the within-community and across-community user characteristic similarity, where each community consists of users who purchase from the same seller (note that users who buy from multiple sellers are considered part of multiple communities). To calculate the similarity within a community, we compute the probability that two members of the community share the same characteristic. In terms of across-community similarity, we randomly partition the communities into two subsets with equal number of communities, and sequentially number the communities in each subset. Then the similarity between the counterparts in the two subsets are computed. Here the similarity is represented by the probability that users share the same characteristics when being randomly selected from the counterparts. The greater the value, the greater the similarity.

Fig. 15 illustrates the proximity of users' cities, provinces and region economic status respectively. From Fig. 15(a) and Fig. 15(b), we conclude that Beidian purchase network is of high physical proximity. Users within the same community have a high likelihood of being physically nearby. For instance, At the city level, 57% communities have within community similarity over 0.6, while the probability of finding two communities whose members share geographical similarity over 0.6 is only 0.004. At the province level, the within group user geographic similarity is even higher. 73.5% communities have within community similarity over 0.6, while the probability of finding communities with across-community similarity over 0.6 is 0.05. As demonstrated by Fig. 15(c), people within the same community are also more alike in their region economic status and are thus more likely to resemble each other in their financial and social status. In fact, 37.6%, 52.7% and 41.4% of all Beidian communities consist exclusively of people sharing the same geographic characteristic on city level, province level and region economic status. Thus, Beidian users demonstrate high level of physical and social proximity, and we argue that this very characteristic is possibly due to the fact that Beidian network is built on homophilous WeChat social network.

**Conversion Rate vs. Proximity** We further show that Beidian's conversion rate is closely linked with user proximity within each seller community. Fig. 16 illustrates the community level conversion rate as a function of within-community user proximity. We can observe that on all three levels - city level (Fig. 16(a)), province level (Fig. 16(b)), and region economic status level (Fig. 16(c)) - the community level conversion rate grows as buyer similarity increases, indicating that more proximate community generally yields higher conversion rate. A possible explanation is that in communities of high level of proximity, sellers and shoppers share many similar characteristics and the sellers understand the needs of their customers better. Thus, in contrast to communities of lower level of proximity, these sellers are more likely to introduce products of interest to community members, which results in higher conversion rate.

### Loyalty

Another intriguing characteristic of Beidian is the loyalty of buyers, which reflects the extent to which buyers are devoted to purchase from the same source. Fig. 17 depicts the distribution of the number of sellers one buyer buys from. We observe that 92.4% of buyers purchased from only one Beidian seller. We have also studied buyers who have at least two WeChat friends acting as sellers (i.e., buyers who visited links sent by at least two sellers), which account for 23.09% of all buyers. Among them, we found 78.2% buyers choose to buy from one seller. Furthermore, we have analyzed the percentage of buyers who carry out repeated purchases at the same seller: while 89.7% of Beidian buyers return to carry out repeated purchases, less than 1% buyers on other platforms do (Gupta and Kim 2007; Zhang et al. 2011). These facts demonstrate that Beidian buyers generally stick to the same seller, who are likely to be of closest relationship to them, and show high degree of loyalty.

To further investigate the relationship between number of purchase sources and purchase numbers, we examine the average number of sellers one buys from, where we group buyers who have made the same number of purchases. The result is presented in Fig. 18. As shown, the average number of purchase sources demonstrates a growing trend as the number of purchase increases. Yet, the increase rate is significantly smaller than purchase number, which demonstrate that buyer loyalty does play an important role in differentiating Beidian from traditional E-Commerce, where buyers often focus on factors other than relationship (e.g. information quality, price, reputation)(Park and Kim 2003) and higher level of trust generally mean participants engage in fewer repeat transactions over same sellers on the 'stranger' marketplace (Bolton, Katok, and Ockenfels 2004; Cai et al. 2014). In contrast, Beidian buyers tend to shop from familiar sellers, probably originate from the trust of strong social ties.

**Conversion Rate vs. Loyalty** Finally, we show the relationship between conversion rate and buyer loyalty. As illustrated in Fig. 19, individual buyer conversion rate decreases as the number of sellers the buyer shops from increases, which indicates that conversion rate positively correlates with buyer loyalty, possibly due to the trust aroused from intimacy.

To sum up, purchase behavior on Beidian shows high conversion rate on site level, community level, individual level and item level. Within communities formed by sellers, users demonstrate the effect of homophily, and higher within-community user similarity correlates with higher conversion rate. Purchase behavior on Beidian also shows high buyer loyalty, where buyers tend to purchase from the same seller(s) over time. Higher user loyalty correlates with higher conversion rate.

## Discussion

So far we have taken a quantitative approach to analyze the fundamental characteristics of recent emerging strong-tie based social commerce sites. As revealed by our analysis, Beidian demonstrates rapid growth and high conversion rate compared to traditional e-commerce sites. We attribute these characteristics as an effect of strong-tie based E-Commerce's unique mechanism.

**Real Life Social Network + E-Commerce.** In the same vein, recent emerging social commerce blends the features of real life social network and E-Commerce. While cascades and proximity have been widely observed and studied in various existing social networks (McPherson, Smith-Lovin, and Cook 2001; Anderson et al. 2015), and the act of word-of-mouth recommendation and viral marketing commonly used in e-commerce sites (Leskovec, Adamic, and Huberman 2007), these factors are not well integrated until the recent surge of social commerce. In Beidian, a number of unconventional scenarios emerge as these factors confluence. For instance, due to the effect of proximity, shoppers and sellers coming from the same community share similar interests and economic status. Thus, sellers are more likely to introduce items of interest to the community than traditional shoppers, potentially leading to high conversion rate.

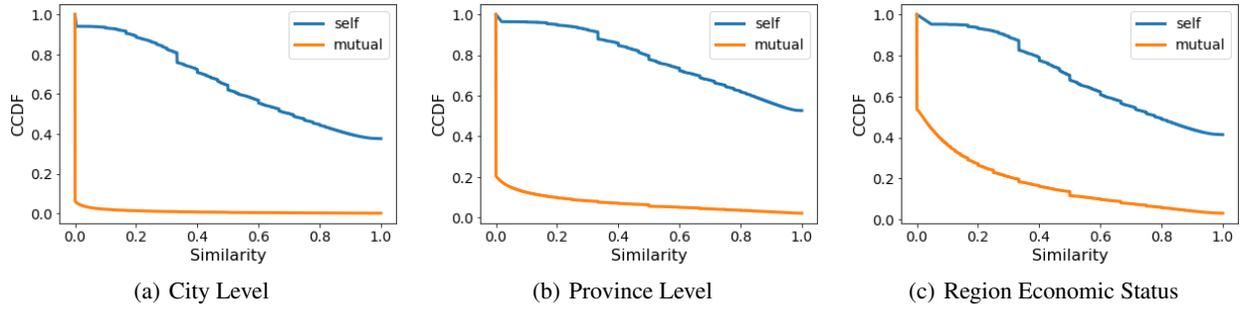

Figure 15: Geographical proximity. Blue curve represents CCDF of within-community geographical similarity, while orange curve represents CCDF of across-community geographical similarity.

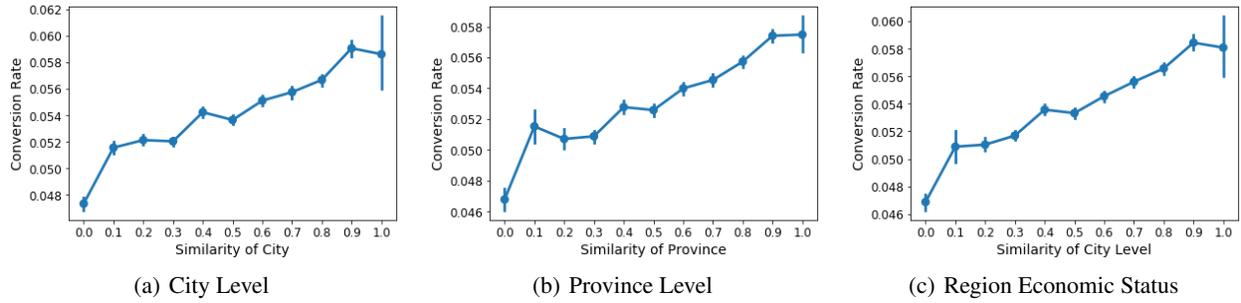

Figure 16: Conversion rate vs. within-community user similarity of social demographic characteristics.

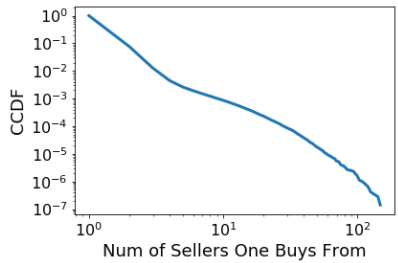

Figure 17: Number of sellers one buys from.

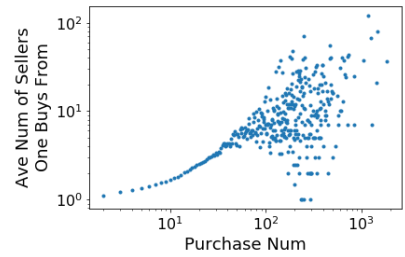

Figure 18: Seller number vs. purchase number.

In essence, strong-tie based commerce makes the first attempt to integrate user economic and social life. It shows how market exchanges in the digital age are not simple individual decisions based on rationalizing and calculating, but also under social influence, particularly from intimate others.

**Trust in Close Ties.** Emerging social commerce also transforms the traditional roles of seller and shopper. In traditional setting, sellers aim at maximizing their profits while shoppers attempt to get the best items at a bargain (Esmaeili, Aryanezhad, and Zeephongsekul 2009). Thus, there exists an inherent opposition between the two roles. Social commerce, however, makes the first attempt to reconcile such conflicts. On Beidian, purchase is made on top of a decentralized and localized network: one buys from his/her friends or relatives in his/her network of close relationships, rather than from certain key opinion leaders, such as big brands. One can even become a seller himself easily if he wants to. Therefore, social commerce creates a network that encourages everyone's participation. As a result, trust between the seller and shopper may establish more organically as social relationships suppress potential opportunistic behavior, which is likely to result in buyer loyalty and higher conversion rate (Becerra and Korgaonkar 2011); meanwhile, trust could influence people's attitude towards sign-up invitation, getting new users more likely to accept the platform, thus indirectly contributes to the massive invitation cascades and fast growth of the platform. We argue that the trust in one's close ties is likely to lay the foundation of social commerce and may be the key reason behind its recent success.

**Economic Gain vs. Toll on Social Tie?** On the other hand, it is important to note that there are potential downsides of the strong-tie based social commerce model. For instance, there could be conflicts between seller economic be-havior and social relationship. If a seller gets more benefit-driven and engage in increasingly expedient behavior, e.g. over advertising, the seller may eventually ruin his/her social tie, which are most likely to be close relationship on WeChat. On a more global scale, misbehavior on such social commerce may also hurt the entire social networking ecosystem (e.g. WeChat), where the commerce heavily relies on. Thus, how to balance the trade-off between economic gain and social relationship, is a question of significance for future research.

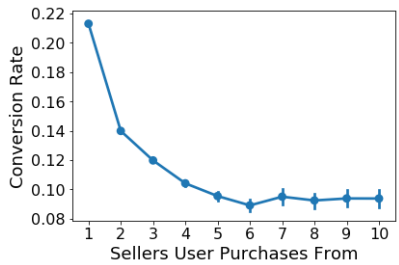

Figure 19: Conversion Rate vs. Loyalty.

**Generalizability, Implications, and Future Work.** Though our work only analyzes the Beidian network, we argue that our findings are generalizable to other strong-tie based social commerce site. In fact, many social commerce sites, such as Yunji[6], operate similarly as Beidian. Take Pinduoduo, the largest and most successful social commerce sites now in China as another example. It acts in a group buying mode, where users share links of items to groups on WeChat and persuade others to join the group buying. The more people joining the group buying, the price of the item will be cheaper. The link sharer in Pinduoduo, however, is in nature a similar role as seller in Beidian, since they are both information sharer within their intimate relationships and are able to introduce products of interest due to the proximity effect.

Our study thus provides important insights for recent emerging strong-tie based social commerce sites as a whole. In this work, we showed how emerging social commerce succeeds and prospers as the consequences of interactions between social closeness and economic logics, where trust of the strong ties lies in the very heart of it. Therefore, for social commerce practitioners, it is always important for to maintain trust on the platform, avoiding them being spoiled by economic activities. For researchers and designers, our work can be considered as a demonstration of the surprising capability of collective intelligence built on seemingly subtle power of ordinary people. It would be of interest to think of the emerging social commerce as a huge 'crowdsourcing information sharing system' embedded in online social networking sites, where sellers introduce items of potential interest to their community members. We envision that there are immense possibilities in turning the existing social network structure into powerful social computing systems that supports complicated goals, in a similar way as intimacy-based social commerce.

As the first quantitative work on strong-tie based social commerce, our study mainly focuses on the measurement and understanding of fundamental characteristics of social commerce platforms. There are still many key research problems yet to be investigated in future work, including user characterization, behavior prediction and recommendation on social commerce (i.e. social recommendation (Lin, Gao, and Li 2018; 2019)). As one of the most revolutionary and exciting trends in recent years, more attention are called for in the data science and computational social science community on the emerging field of social commerce.

# Conclusion

Leveraging the full-scale data of Beidian, one of Chinese fastest growing WeChat based social commerce, we presented the first quantitative measurement on the network structure and dynamics of recently emerging real life social network based E-Commerce. In comparison with traditional E-Commerce sites, Beidian is decentralized, localized, fast growing and demonstrates high conversion rate. We identified invitation cascade as the main driver of Beidian's fast growth, where Beidian shows cascade patterns that are much deeper and larger in size than prior networks. We attributed Beidian's high conversion rate as a likely consequence of network proximity and user loyalty.

As one of the most revolutionary and exciting trends in recent years, social commerce presents novel application scenarios, challenges and possibilities for computer science, network science, sociology, psychology, marketing and organization science. Our work offers the first glimpse into the potentials of this new research field.

# Acknowledgement

The authors thank Daniel A. McFarland, Michael S. Bernstein, Bas Hofstra and Vivek Kulkarni for helpful discussions and suggestions. We also thank the associate editor and reviewers for their valuable comments. This work was supported in part by The National Key Research and Development Program of China under grant SQ2018YFB180012, the National Nature Science Foundation of China under 61861136003, 61621091 and 61673237, Beijing National Research Center for Information Science and Technology under 20031887521, and research fund of Tsinghua University - Tencent Joint Laboratory for Internet Innovation Technology. Hancheng Cao was supported by The James D. Plummer Graduate Fellowship - a Stanford University School of Engineering (SoE) Fellowship.

---

[6]https://www.yunjiweidian.com/